\documentclass[twocolumn,aps,prl,10pt,superscriptaddress,letterpaper,citeautoscript,showpacs]{revtex4-1}

\usepackage{amsfonts,amsmath,amssymb}

\usepackage{graphicx}

\def\usedvipdfmx{0} % % 1 for dvipdfmx, 0 for pdflatex
\def\inclfig{1} % % 1 for including figure, 0 for excluding
\def\inclrefttl{0} % % 1 for including reference titles, 0 for excluding them

\graphicspath{{./FigureMain/},{./FigureSuppl/}}

\if \usedvipdfmx 1
\else
\fi

\if \inclrefttl 1
  \def\articletitle#1{{\it #1},}
\else
  \newcommand{\articletitle}[1]{}
\fi

\usepackage{color}
\usepackage{ulem}

\newcommand{\ket}[1]{ |{#1} \rangle}
\newcommand{\bra}[1]{ \langle {#1}|}
\newcommand{\sq}[1][]{\hat{S}_{\mathrm{#1}}(\gamma )}

\newcommand{\nature}{ Nature }
\newcommand{\natphoton}{ Nat.\ Photon.\ }
\newcommand{\science}{ Science }
\renewcommand{\prl}{ Phys.\ Rev.\ Lett.\ }
\newcommand{\prx}{ Phys.\ Rev.\ X }
\renewcommand{\pra}{ Phys.\ Rev.\ A }
\newcommand{\prev}{ Phys.\ Rev.\ }
\renewcommand{\rmp}{ Rev.\ Mod.\ Phys.\ }
\newcommand{\optexp}{ Opt.\ Exp.\ }

\newcommand{\job}{ J.\ Opt.\ B }
\newcommand{\epl}{ Europhys.\ Lett.\ }

\newcommand{\UTokyo}{
Department of Applied Physics, School of Engineering, \\
The University of Tokyo, 7-3-1 Hongo, Bunkyo-ku, Tokyo 113-8656, Japan}
\newcommand{\UPalacky}{
Department of Optics, Palack\'{y} University,
17. listopadu 1192/12, 772 07 Olomouc, Czech Republic}
\newcommand{\UMainz}{
Institute of Physics, Staudingerweg 7,
Johannes Gutenberg-Universit\"{a}t Mainz, 55099 Mainz, Germany}

\begin{document}

\title{Exploring a new regime for processing optical qubits:\\
squeezing and unsqueezing single photons}

\author{Yoshichika Miwa}
\affiliation{\UTokyo}
\author{Jun-ichi Yoshikawa}
\affiliation{\UTokyo}
\author{Noriaki Iwata}
\affiliation{\UTokyo}
\author{Mamoru Endo}
\affiliation{\UTokyo}
\author{Petr Marek}
\affiliation{\UPalacky}
\author{Radim Filip}
\affiliation{\UPalacky}
\author{Peter van Loock}
\affiliation{\UMainz}
\author{Akira Furusawa}
\email{akiraf@ap.t.u-tokyo.ac.jp}
\affiliation{\UTokyo}

\date{\today}

\begin{abstract}
We implement the squeezing operation as a genuine quantum gate, deterministically and reversibly
acting `online' upon an input state no longer restricted to the set of Gaussian states.
More specifically, by applying an efficient and robust squeezing operation for the first time to
non-Gaussian states, we demonstrate a two-way conversion between a particle-like single-photon state and a wave-like superposition of coherent states.
Our squeezing gate is reliable enough to preserve the negativities of the corresponding Wigner functions.
This demonstration represents an important and necessary step towards hybridizing discrete and continuous
quantum protocols.
\end{abstract}

% % 03.67.-a : Quantum information
% % 03.67.Hk : Quantum communication
% % 42.50.Dv : Quantum state engineering and measurements
% % 42.50.Ex : Optical implementations of quantum information processing and transfer

\maketitle

From a fundamental point of view, quantum states of light can behave in a complementary fashion, showing both particle-like and wave-like behavior. With regards to an application such as quantum computing, an important
proposal for universally processing photonic qubits \cite{KLM} makes use of quantum particle detections
(i.e., photon counting) and quantum wave evolutions (i.e., quantum interferences through
passive, energy-preserving linear optical circuits).
In this scheme, however, the required ancilla states consist of many highly entangled photons and thus
are out of reach of current experimental capabilities. It is therefore reasonable to extend the toolbox of optical
quantum operations in order to reduce the cost of the necessary quantum resources.
For instance, apart from discrete `click by click' measurements that rely on the particle-like nature of light \cite{Ionicioiu11}, quantum light fields may be detected more naturally via continuous
phase-space measurements exploiting their wave-like features \cite{Glauber63}.
In fact, two recent continuous-variable teleportation experiments on non-Gaussian input states,
using Gaussian entanglement and Gaussian homodyne measurements,
demonstrate that not only a wave-like coherent-state superposition (CSS) \cite{CatTeleportation}, but also a particle-like photonic qubit \cite{DRQubitTeleportation} can be transferred efficiently and reliably,
preserving the negativity of the Wigner function and exceeding the classical fidelity limits, respectively.

In order to further investigate the potential of such {\it hybrid} schemes \cite{hybridbook}, which simultaneously exploit discrete and continuous techniques for encoding, measuring, and processing quantum states of light, it is desirable, besides Gaussian measurements and resource states,
to also add Gaussian gate operations to the set of possible
optical elements for processing photonic qubits.
Indeed, one feasible regime of single-photon operations has still remained
unexplored: Gaussian operations including {\it active squeezers}, still
linearly transforming the mode operators, but no longer preserving energy.

% There are already prominent proposals for quantum computation \cite{KLM} that employ linear optics in order to process photonic qubits, making use of both quantum particle correlations and quantum wave interference. Everything needed for this are passive, photon-number-preserving beam splitting operations and highly entangled multi-mode, multi-photon states of sufficiently many photons. In this approach, the online operations, i.e., the gate operations that sequentially act on some encoded quantum information, are passive, photon-number-preserving beam splitting and phase shifting operations, almost as simple and practical as they can possibly be. However, unfortunately, the offline preparation of the required ancilla states, which consist of many photons, is out of reach of current experimental capabilities. Therefore, it seems reasonable to gradually extend the online toolbox in order to reduce the cost of the offline resources. Nonlinear optical operations that transform the optical mode operators in a nonlinear fashion, though being sufficient for universal quantum processing, are still hard to obtain experimentally. However, there is one regime below this which is much more feasible and could be explored: that of Gaussian operations including active squeezers, still linearly transforming the mode operators, but no longer preserving photon numbers. What we demonstrate here can be considered as this addition of active elements to the online toolbox.

The squeezing operation has been traditionally associated with continuous-variable quantum optics and information \cite{HistoricalSqueezingRef,RMP_Braunstein,RMP_Weedbrook}.
In fact, it can be considered the essential, elementary operation of this area, as it is a necessary component of all Gaussian gates \cite{irreducible} and even some non-Gaussian gates require it for their implementation \cite{x3gate}.
The construction of a genuine squeezing {\it gate} is fundamentally different from simply preparing a particular squeezed {\it state}. This basic difference was previously addressed in an experiment employing a measurement-based protocol \cite{Yoshikawa07.pra}.
Later, this scheme was extended to more advanced Gaussian gates by using the squeezers \cite{Yoshikawa07.pra} as their fundamental building blocks: a quantum nondemolition sum gate \cite{Yoshikawa08.prl}, which may be understood as a continuous-variable version of the controlled-NOT gate for qubits, and a reversible phase-insensitive amplification (two-mode squeezing) gate \cite{Yoshikawa11.pra}, which also functions as an approximate cloner for coherent-state inputs.
However, in all these previous demonstrations, only Gaussian states have been used for the inputs.
One reason for this is of technical nature:
the bandwidth of the squeezing gate is typically very narrow and there has been no way so far to generate highly non-classical, non-Gaussian states in such a narrow frequency band.
Moreover, these non-Gaussian states tend to be extremely sensitive to losses, and thus, coupling them directly into an optical parametric oscillator will easily erase any signature of their strong nonclassicality such as the negativity of the Wigner function.
Our demonstration here was made possible by introducing a recent technique for bandwidth broadening as well as a mechanism for increased loss robustness to the squeezing gate \cite{CatTeleportation}.

We experimentally demonstrate for the first time a deterministic squeezing gate that operates on non-Gaussian input states.
In particular, as what we believe to be a nice illustration, we use a particle-like single-photon state as the input state of the squeezing gate.
The resulting output state then is a wave-like CSS.
Since single-mode squeezing corresponds to a unitary, noiseless amplification process along a certain phase-space direction, our single-photon squeezer can be also interpreted
as a phase-sensitive amplifier acting on an optical field mode in its first excited state (for a more detailed discussion of this interpretation, see the Supplemental Material \cite{Supplemental}).

Furthermore, we also demonstrate the inverse operation of the squeezer, where a wave-like CSS is converted into a particle-like single-photon state.
From a more fundamental point of view, what we demonstrate here can be considered an unconditional and reversible two-way conversion between a single quantum particle and a non-classical, continuous wave.
Unlike the previous probabilistic conversion from a photon number state to a CSS \cite{CSSfromPhotonEx}, the squeezing gate deterministically and reversibly transforms a single photon into a CSS.
The CSS, $\ket{\alpha} - \ket{-\alpha}$, where $\ket{\alpha}$ is a coherent state \cite{Glauber63}, is a highly non-classical quantum state sufficient for universal quantum computation \cite{CSS_comp}.
It is worth noting that an all-optical, high-purity, almost-on-demand single-photon source was reported recently \cite{Yoshikawa13}, while no such source has ever been demonstrated for a CSS state.
Therefore, our unconditional conversion between these two types of states means that, in principle, all such quantum resources, including CSS states, are now available nearly on demand.

We believe that this experiment paves the way for quantum applications that combine discrete-particle and continuous-wave protocols in a so-called hybrid fashion.
The squeezing-gate operation when acting on non-Gaussian states has also a number of direct applications, such as quantum state discrimination of optical coherent-state qubits \cite{Takeoka}, Gaussian optimization in non-Gaussian state preparation \cite{Menzies09.pra}, improved quantum state transmission through a lossy channel \cite{Filip13.pra}, and preprocessing before a light-matter coupling for an efficient quantum memory interface \cite{Interface}.
Recently, it has been also realized that squeezing is an extremely useful tool for manipulating and measuring individual photons, for instance, in squeezing-enhanced Bell measurements of optical qubits \cite{Zaidi} or in squeezing-enhanced entanglement distillation protocols where optical Gaussian states are locally transformed into qubit Bell pairs \cite{AkiraDistillationExperiment,TheorySqueezingHelps}.

\begin{figure}
\centering
\if \inclfig 1
\includegraphics[clip]{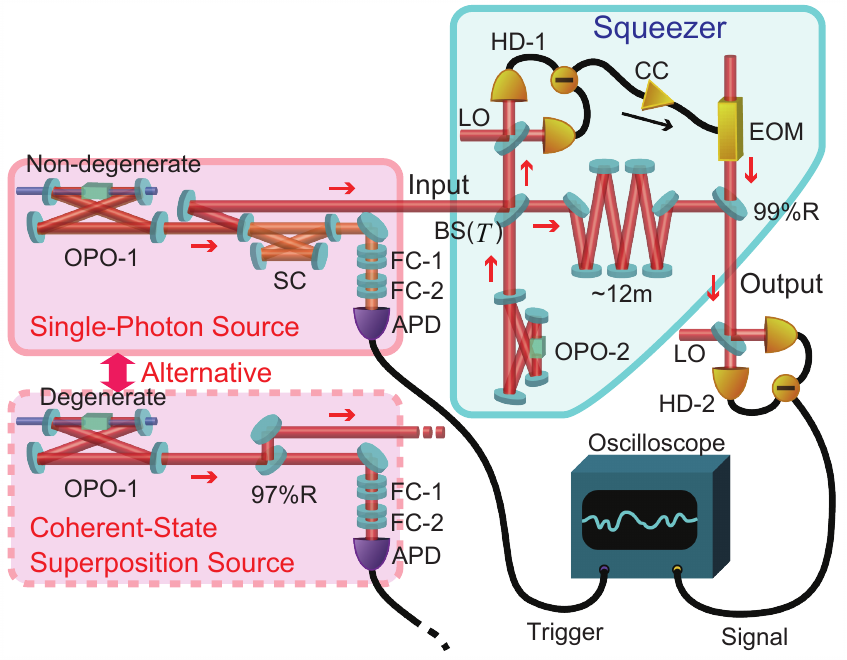}
\fi
\caption{
(Color Online)
Experimental setup.
BS($T$): beam splitter with transmittance $T$ determining the degree of squeezing,
OPO: optical parametric oscillator,
SC: separating cavity,
FC: filter cavity,
APD: avalanche photo diode,
HD: homodyne detector,
LO: optical local oscillator,
EOM: electro-optic modulator,
CC: classical channel,
R: reflectivity.
An optical delay line of about $12$~m, traveling in free space, is used
to match the propagation times of the two signals,
one of which gets converted to an electrical signal and back,
while the other one remains optical throughout.
}\label{fig:setup}
\end{figure}

% $\hbar = 1$

The schematic of our experimental setup is shown in Fig.~\ref{fig:setup}.
It consists of two parts:
a source of non-classical states and an unconditional squeezer.
For the former, via a small variation of the setup, we can choose the non-classical states to be either a single photon \cite{JonasPhoton.optex} or a CSS \cite{WakuiCat.optex}.
The states will always emerge randomly in time, however, from a photon ``click'' at the avalanche photodiode (APD), we know whenever a state arrives.
These ``heralded'' non-classical states are localized in time around the detections of correlated photons \cite{JonasPhoton.optex, WakuiCat.optex}.
Therefore, our unconditional squeezer must have enough bandwidth to be applicable in the corresponding short time slots \cite{CatTeleportation}.
We extended our previous measurement-based squeezer \cite{Yoshikawa07.pra} to meet this requirement.
This squeezer avoids direct coupling of fragile input states to nonlinear optical media, which typically involves large optical losses.
Instead, an ancillary squeezed state is utilized as a resource of nonlinearity \cite{Filip05.pra} (see the Supplemental Material \cite{Supplemental}).
In this scheme, the higher the squeezing level of the ancilla state becomes, the more closely the squeezing operation resembles a unitary, completely reversible squeezing gate.
Although our squeezer is assisted by homodyne detection on the ancilla beam, the non-classical signal state is never directly measured (see the quantum eraser~\cite{Andersen04.prl}) when the squeezer is applied.

\begin{figure*}
\centering
\if \inclfig 1
\includegraphics[clip]{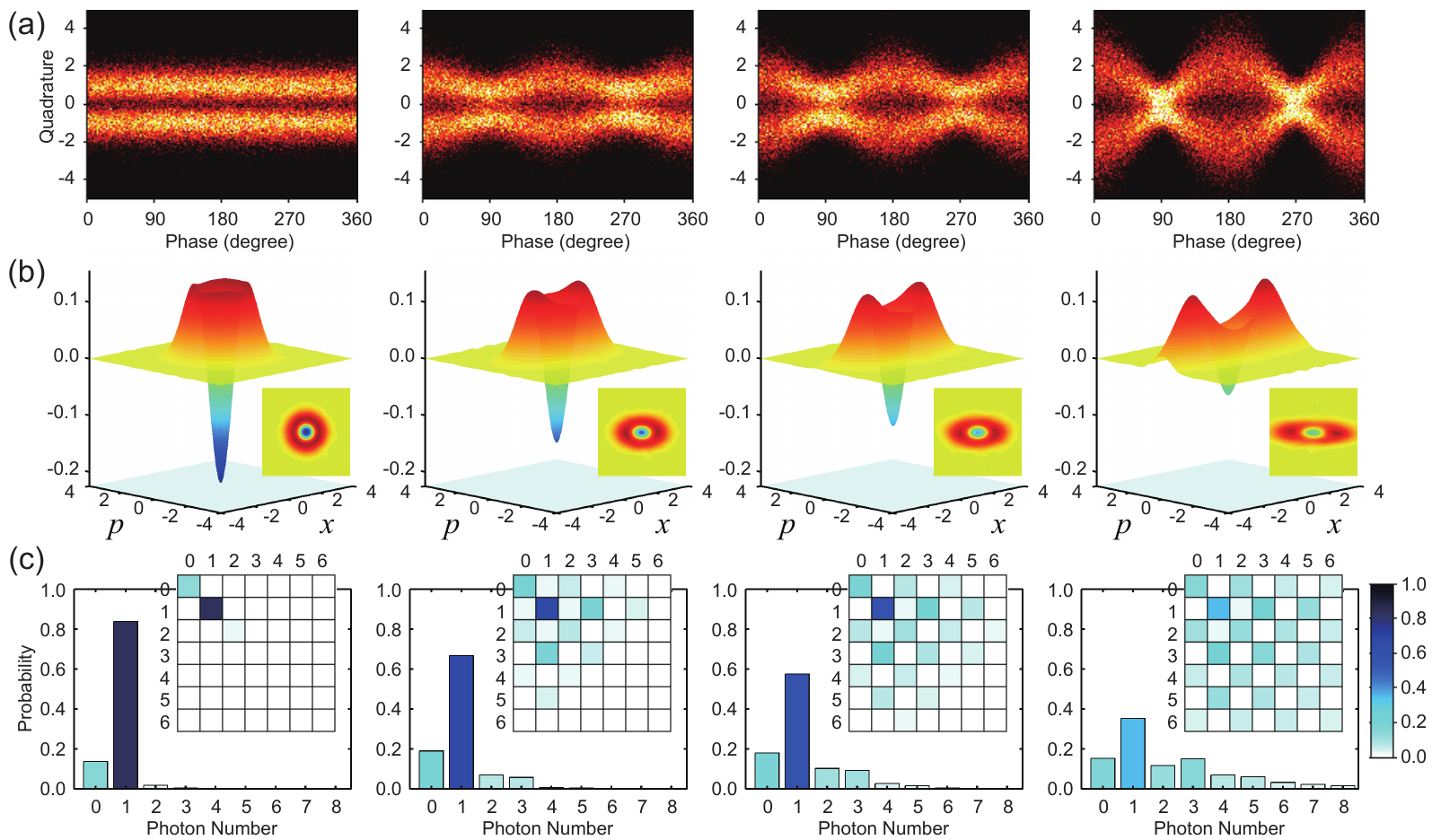}
\fi
\caption{
(Color Online)
Experimental quantum states for the conversion from particle to wave.
The leftmost column shows the input single-photon state, while the other three columns show the output states for a squeezing parameter $\gamma$ of $0.26$, $0.37$, and $0.67$, from left to right.
(a) Quadrature distributions over a period.
(b) Wigner functions.
(c) Photon number distributions and photon number representation of density matrices.
The minimum value of $-0.22$ for the input Wigner function becomes, respectively, $-0.15$, $-0.12$, and $-0.06$, after the conversion.
}\label{fig:results1toCSS}
\end{figure*}

In order to verify the conversions, we perform quantum homodyne tomography
on input and output states \cite{Lvovsky09.rmp, Lvovsky04}.
Recall that wave and particle properties in quantum optics are formally connected via a pair of annihilation and creation operators for photons, $\hat{a}$ and $\hat{a}^\dagger$, respectively.
These non-Hermitian operators are the quantized versions of the complex and complex conjugate amplitudes of an optical field mode, satisfying the bosonic commutation relation $[\hat{a},\hat{a}^\dagger] = 1$.
Similarly, the quadrature operators,
$\hat{x} = (\hat{a} + \hat{a}^\dagger) / \sqrt{2}$ and $\hat{p} = (\hat{a} - \hat{a}^\dagger) / i\sqrt{2}$,
correspond to the quantized real and imaginary parts of the optical complex amplitudes (up to a factor $\sqrt{2}$),
where $[\hat{x},\hat{p}] = i$.
Through homodyne detection, the quadrature $\hat{x}(\theta)$ can be measured, which gives an Hermitian part of the operator $\hat{a}e^{-i\theta}$;
$\theta = 0$ and $\theta = \pi / 2$ then correspond to $\hat{x}$ and $\hat{p}$, respectively.

\begin{figure}
\centering
\if \inclfig 1
\includegraphics[clip]{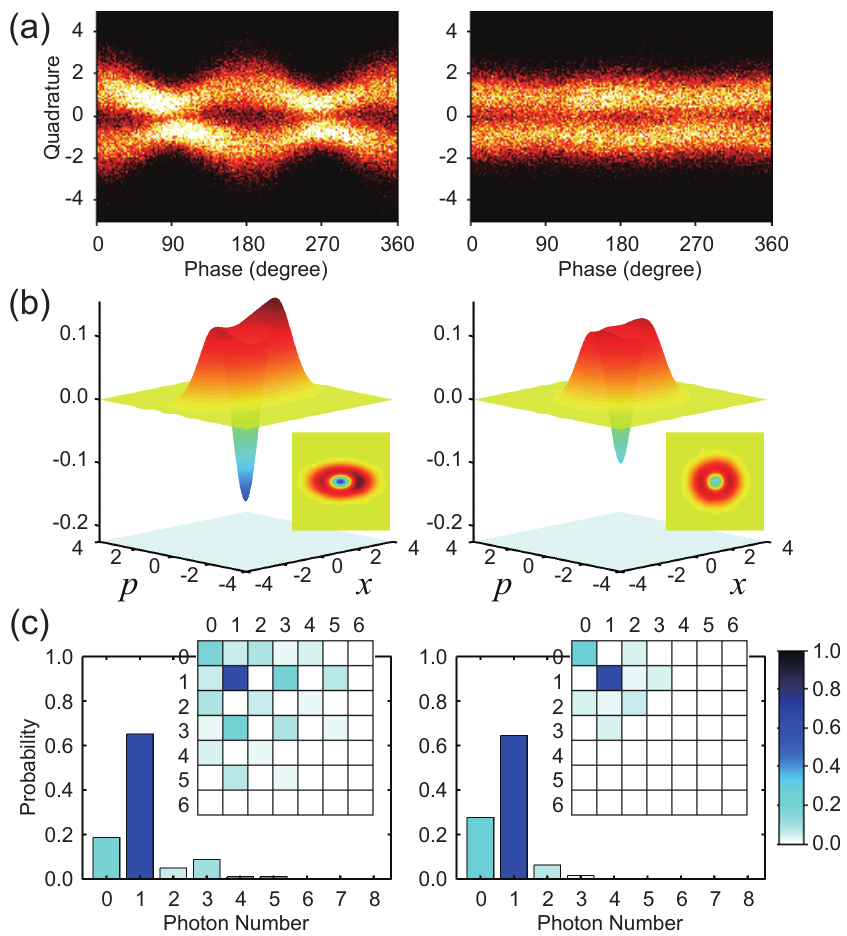}
\fi
\caption{
(Color Online)
Experimental quantum states for the conversion from wave to particle.
The left column shows the input coherent-state superposition, while the right column shows the output state for a squeezing parameter $\gamma$ of $-0.26$.
(a) Quadrature distributions over a period.
(b) Wigner functions.
(c) Photon number distributions and photon number representation of density matrices.
The minimum value of $-0.16$ for the input Wigner function becomes $-0.10$ after the conversion.
}\label{fig:resultsCSSto1}
\end{figure}

The experimental results for converting single-photon states into several CSSs are shown in Fig.~\ref{fig:results1toCSS}, and those for the reciprocal conversion are given in Fig.~\ref{fig:resultsCSSto1}.
The top panels show the phase dependence of quadrature distributions obtained by a series of homodyne measurements.
From these, Wigner functions and photon-number density matrices are calculated, as shown in the lower panels.
In Fig.~\ref{fig:results1toCSS}, the leftmost column shows the input single-photon state, while the three right columns show the output CSSs for three different squeezing levels.
Similarly, in Fig.~\ref{fig:resultsCSSto1}, the left column shows the input CSS, and the right column shows the output single-photon state.

We shall first discuss the quadrature distributions (top panels).
The Fock state $\ket{1}$ of a single photon, which is a typical carrier of discrete-variable quantum information, is a highly non-classical energy eigenstate of a quantized oscillator with a totally undetermined phase.
The phase-insensitivity of the quadrature distribution is a characteristic of a single-photon state, as can be seen in the leftmost panel of Fig.~\ref{fig:results1toCSS}(a) and the right panel of Fig.~\ref{fig:resultsCSSto1}(a).
On the other hand, any coherent state is an eigenstate of the annihilation operator,
$\hat{a}\ket{\alpha} = \alpha \ket{\alpha}$.
This corresponds to a sinusoidal wave with mean complex amplitude $\alpha$ and minimal quantum noise \cite{Glauber63}.
By superimposing two coherent states, $\ket{\alpha} - \ket{-\alpha}$, the quadrature distribution corresponds to two sinusoidal waveforms with quantum interference at each intersection, like in the three right panels of Fig.~\ref{fig:results1toCSS}(a) and the left panel of Fig.~\ref{fig:resultsCSSto1}(a).
This quantum interference is a witness for a genuine quantum superposition of $|\alpha\rangle$ and $|-\alpha\rangle$ and it would never occur for a stochastic mixture of coherent states.

The conversion is achieved by means of a squeezing operation,
$\sq = e^{\gamma (\hat{a}^{\dagger 2} - \hat{a}^2)/2}$,
where $\gamma \in \mathbb{R}$ quantifies the amount of squeezing.
In Fig.~\ref{fig:results1toCSS}(a), the phase-dependent oscillations increase with larger squeezing.
Three different squeezing levels $\gamma = 0.26$, $0.37$, and $0.67$ are demonstrated, resulting in three different amplitudes of CSSs, $\alpha = 0.91$, $1.10$, and $1.64$, respectively. % theoretical amplitudes
The gap at the intersection of the waves becomes less pronounced for larger $\gamma$ because of the finite squeezing of the ancilla mode (about $7$~dB relative to shot noise).
In an opposite manner, in Fig.~\ref{fig:resultsCSSto1}(a), a phase-dependent oscillation is canceled by squeezing, resulting in a phase-independent distribution with a gap, like for a single-photon state.
The input CSS with $\alpha = 0.97$ is converted by squeezing with $\gamma = -0.26$.

In the corresponding Wigner functions (middle panels),
where detector inefficiencies and losses are not corrected,
the non-classicality of the input and output states becomes manifest in negative values.
These Wigner functions are converted from rotationally symmetric to asymmetric [Fig.~\ref{fig:results1toCSS}(b)] and from asymmetric to symmetric [Fig.~\ref{fig:resultsCSSto1}(b)], while preserving their large negative values at the phase-space origin.
In Fig.~\ref{fig:results1toCSS}(b), the minimum value of $-0.22$ at the input becomes, respectively, $-0.15$, $-0.12$, and $-0.06$, at the output.
In Fig.~\ref{fig:resultsCSSto1}(b), $-0.16$ at the input becomes $-0.10$ at the output.

The density matrices (bottom panels) represent the particle picture.
In the particle picture, the effect of the squeezing is an infinite superposition of photons added and subtracted in multiples of two.
As a result, squeezing leads to a superposition of even photon number states when applied to the vacuum and to a superposition of odd photon number states when applied to the single-photon state.
Being in such an odd-number superposition is also a distinct feature of the target CSS,
\begin{align}
\ket{\alpha} - \ket{-\alpha} \, \propto \, \ket{1} + \dfrac{\alpha^2}{\sqrt{6}} \ket{3} + ...\,,
\end{align}
and this is exactly the reason why squeezing achieves the desired conversion \cite{CSSfromPhoton}.

The diagonal elements of the density matrices represent photon number distributions, while the off-diagonal elements correspond to superpositions of $\ket{1}$ and $\ket{3}$.
The input single-photon state in Fig.~\ref{fig:results1toCSS}(c) has a dominant single-photon component of $84\%$ (without any corrections), while the input CSS in Fig.~\ref{fig:resultsCSSto1}(c) has dominating one- and three-photon components compared to the \mbox{zero-,} \mbox{two-,} and four-photon terms.
This also holds for the off-diagonal interference terms such as $\ket{1}\bra{3}$.
Two-photon creations and annihilations are revealed by an increase [Fig.~\ref{fig:results1toCSS}(c)] and a decrease [Fig.~\ref{fig:resultsCSSto1}(c)] of the three-photon components, respectively.

In order to quantitatively assess the experimental conversion processes, besides reconstructing the Wigner functions and density matrices of the input and output states, we used two additional figures of merit.
These are specifically designed to reveal either the most distinct features of the particle-to-wave transition
or that of the converse, wave-to-particle transition \cite{anti-correlation} (for details, see the Supplemental Material \cite{Supplemental}).

From an experimental point of view, it has been considered notoriously hard to apply a quantum optical squeezing operation upon more exotic, non-Gaussian quantum states such as discrete-variable single-photon states.
In our experiment we have succeeded in this difficult task.
By demonstrating the efficient and deterministic squeezing and unsqueezing of a single photon, we have opened an entirely new optical toolbox for future quantum information applications.
In principle, such a unitary, phase-sensitive amplifier (and attenuator) will allow for making use of the entire Fock space when processing single photons, which may help to construct quantum gates and error correction codes for logical qubits.
Using our universal and reversible low-loss broadband squeezer, we have for the first time access to a complete set of deterministic Gaussian operations applicable to non-classical, non-Gaussian states.
These expand the toolbox for hybrid quantum information processing \cite{hybridbook}, and therefore our result will directly lead to applications in this area.
At the same time, besides providing a completely new class of optical quantum processors, our experiment bridges two quantum-mechanically distinct regimes: that of particle-like quantum states such as single photons with that of more wave-like states such as coherent-state superpositions.

This work was partly supported by
PDIS, GIA, G-COE, and APSA commissioned by the MEXT of Japan,
FIRST initiated by the CSTP of Japan, SCOPE program of the MIC of Japan, and, %ASCR-JSPS.
JSPS and ASCR under the Japan-Czech Republic Research Cooperative Program.
R.F.\ acknowledges projects P205/12/0577 of GA \v CR.
P.M.\ acknowledges projects GPP205/10/P319 of GA \v CR and LH13248 of Czech Ministry of Education.
P.v.L.\ acknowledges support from the BMBF in Germany (QuOReP and HIPERCOM).

\makeatletter
% \close@column@grid
\balancelastpage@sw{%
  \onecolumngrid
 }{}%
\newpage
\twocolumngrid
\makeatother

% \clearpage

% \makeatletter
% \close@column@grid
% \balancelastpage@sw{%
%   \onecolumngrid
%  }{}%
% \twocolumngrid
% \makeatother

% \clearpage

\begin{center}\Large\bf Supplemental Material\end{center}

\section{Supplemental Discussion}

When squeezed, the single photon ceases to be a single quantum particle and it becomes a superposition in an infinite-dimensional Hilbert space.
As such, it gains many useful abilities exploited in the various quantum protocols \cite{hybridbook,KLM,HistoricalSqueezingRef,RMP_Braunstein,irreducible,CSS_comp}, but its description and, consequently, analysis becomes more involved.
However, we can greatly benefit from the well-known similarity between squeezed single photons and superposed coherent states \cite{CSSfromPhoton}.
If we are able to rigorously confirm that a single photon was transformed into a high-quality CSS and that a CSS was transformed back into a high-quality single photon, we can safely claim that the squeezing operation works as intended and that it should have its potential place in future applications.

%\textbf{Conversion in Wave Picture} --- In the wave picture, squeezing acts on the single photon state as a phase-sensitive amplifier, amplifying one quadrature ($\hat{x}$) while noiselessly damping (squeezing) the other one ($\hat{p}$). Now notice that, due to duality, the single-photon state can be recast as a {\em  continuous} superposition of CSSs with all possible phases, $\ket{1}\propto \int (\ket{\alpha e^{i\phi}} - \ket{-\alpha e^{i\phi}})d\phi$. The squeezing then drives the coefficients of the CSSs towards the states $\ket{\alpha}$ and $\ket{-\alpha}$, and we obtain a {\em discrete} superposition, $\sq \ket{1} \approx \ket{\alpha}-\ket{-\alpha}$, especially when $|\alpha| < 1.2$, with an approximation error less than $1\%$~\cite{CSSfromPhoton}. By inverting the unitary squeezing, the CSSs are converted back into the single-photon state, $\sqm \left( \ket{\alpha}-\ket{-\alpha} \right) \propto  \ket{1}$, regaining their phase independence within the continuous superposition forming $\ket{1}$.

\subsection{Conversion Criteria}

In order to quantitatively assess the transition between the single photon and the CSS, we employ the following figures of merit.
First, note that the ideal initial and target states, the Fock state
$|1\rangle$ and the coherent-state superposition
$|{\rm CSS_{id}}(\alpha)\rangle\propto |\alpha\rangle - |-\alpha\rangle
\propto |1\rangle + \alpha^2 |3\rangle/\sqrt{6}$, become identical
when $\alpha \rightarrow 0$. Hence the usual measure of fidelity between the
experimentally generated states  and an ideal state $|\psi\rangle$,
$F=\langle\psi|\rho|\psi\rangle$, is not a useful figure of merit. For example,
the overlap between the experimental $\rho$ and a target CSS,
$F(\alpha) =  \langle {\rm CSS_{id}}(\alpha)|\rho|{\rm CSS_{id}}(\alpha)\rangle$
would optimally attain a value close to the maximal value of unity.
This can be achieved when the state $\rho$
approaches an ideal CSS with amplitude $\alpha$, but also
when $\rho$ is close to $|1\rangle$ for $\alpha \rightarrow 0$.
In order to avoid such a confusion between our particle-like
and wave-like states and to gain a better insight into each generated state's
properties, we divide the fidelity into two separate measures:
the coherent-state distinguishability factor, $D(\beta)=\left(\langle\beta|\rho|\beta\rangle+\langle-\beta|\rho|-\beta\rangle\right)/2$,
and the coherent-state interference factor, $V(\beta)=\left(\langle\beta|\rho|-\beta\rangle+\langle-\beta|\rho|\beta\rangle\right)/2$.

%below taken from main text:
%%%%%%%%%%%%%%%%%%%%%
%%%%%%%%%%%%%%%%%%%%%
%One is most appropriate to verify the particle-to-wave transition:
%$D(\alpha) = (\langle \alpha|\rho|\alpha\rangle + \langle-\alpha|\rho|-\alpha\rangle)/2$,
%describing the appearance of the dominant coherent states in the CSS
%and their distinguishability, together with the interference factor
%$V(\alpha) = (\langle \alpha|\rho|-\alpha\rangle + \langle-\alpha|\rho|\alpha\rangle)/2$,
%serving as a witness for those dominant coherent states
%being in a superposition rather than a mixture.
%Squeezing causes the maximum of $D(\alpha)$ to move towards higher values of $\alpha$,
%thus indicating the growth of the CSS.
%At the same time, $V(\alpha)$ remains in the range of values
%confirming nonclassicality of the CSS.
%For the wave-to-particle transition, since any loss of phase properties
%in $D(\alpha)$ and $V(\alpha)$ may as well be explained by an incoherent dephasing,
%we use instead the anti-correlation effect \cite{anti-correlation}
%described by parameter $A$.
%Through squeezing, the indivisibility of a single photon is regained in our experiment.
%%%%%%%%%%%%%%%%%%%%%%
%%%%%%%%%%%%%%%%%%%%%%

Here, $D(\beta)$ describes the average overlap of the investigated state $\rho$ with an independent pair of coherent states with amplitudes $\pm \beta$, which form the ideal superposition. When $\beta=0$, this overlap vanishes for our odd-number CSS, while it would become unity for an even-number CSS, $|\alpha\rangle + |-\alpha\rangle$.
In either case, as well as for an incoherent mixture of coherent states $|\pm \alpha\rangle$,
$D(\beta)$ has, ideally, two separated symmetric peaks with maxima approaching $0.5$ when $|\alpha|> 2$.
On the other hand, the single-photon state $|1\rangle$ has a maximal distinguishability
$D_1^{max}= 0.37$ at $\beta \approx 1$. An additional feature separating the CSS from the single-photon state is
that the distinguishability of $|1\rangle$ is phase-insensitive, $D_1(\beta e^{i\phi}) \equiv D_1(\beta)$, whereas for both the ideal and the experimentally generated CSSs, it is not, having a maximum at $\phi= 0$.

Besides two separated coherent-state peaks in phase space, a CSS is characterized by interference between the coherent states.
For the Wigner function, this interference gives a negativity at the phase-space origin. However, a single-photon state exhibits the same feature, and if we want to distinguish a CSS from a single-photon state, we must evaluate the difference in this interference.
For this purpose, we introduce the interference factor $V(\beta)$, %which is the counterpart to distinguishability for the interference pattern.
which is always positive for a balanced mixture of coherent states and always negative for both the single-photon and the CSS $|\alpha\rangle-|-\alpha\rangle$. This makes it useful for demonstrating nonclassical interference effects.
However, this figure of merit achieves more. While it has a clear minimum for the single-photon state, $V_1^{min}=-D_1^{max}$ at $\beta = 0.97$, for the ideal CSS, its extremal value depends on the amplitude $\alpha$ and this value monotonically decreases to $-0.5$ as $\alpha$ increases. Although both for the ideal single-photon state and the ideal CSS, the distinguishability and the interference are related by $D(\beta) = -V(\beta)$, in the presence of realistic noise, this symmetry relation is broken and we benefit from treating the two quantities independently.

Two benchmarks can be derived to demonstrate that we observe nontrivial interference effects in comparison to the classical theory. The first benchmark represents the interference achievable purely by a mixture of coherent states. Passing it is a first witness of non-classical interference. Passing the second benchmark, obtained through optimization over all possible mixtures of Gaussian states, then serves as a witness of higher-order non-classical interference. Once again we stress that these criteria are different from simply observing negativities,
which is always a witness of non-classicality, but does not discriminate between a CSS and a single-photon state
like our interference factor $V(\beta)$ does. In order to verify the single photon-to-CSS transition in our
experiment, we employ $D(\beta)$ in order to describe the emergence of the two coherent-state peaks in the CSS.
At the same time, $V(\beta)$ enables us to rule out incoherent mixtures and verify the presence
of non-classical interference.

For the CSS-to-single photon transition, we introduce an additional figure of merit, since
any loss of phase properties in $D(\beta)$ and $V(\beta)$ (as a possible indication
for a successful conversion into a single-photon state) may as well be
explained by an incoherent dephasing.
In order to describe this reverse transition from the CSS back to the single-photon state, we take advantage of an important feature of the ideal single-photon state --- it is indivisible at a beam splitter. This property, which has been essential in many single-photon experiments, can be evaluated through the so-called anti-correlation factor $A$ (Ref.~\cite{anti-correlation}).
This operational measure is obtained assuming that the state in question is split at a balanced beam splitter and the two output modes are measured by realistic single-photon detectors. The anti-correlation parameter is then $A = p_c/p_s^2$, where $p_c$ is the probability for both detectors registering a photon, while $p_s$ is the probability for only one detector registering a photon (independent of the other detector). For a mixture of a single-photon and a vacuum state,  $p|1\rangle\langle 1|+(1-p)|0\rangle\langle 0|$, the anti-correlation parameter is $A=0$, while classical light always has $A\ge 1$. As the amplitude of the ideal odd CSS increases, $A=1-[1-2\cosh\left(|\alpha|^2/2\right)]^{-2}$ grows from zero, approaching unity. For the ideal squeezed single-photon state, $A$ increases monotonously with the squeezing and for $|\gamma| > 0.66$, $A$ becomes larger than unity, thus confirming the loss of the single-photon character.

\subsection{Analysis of the experiment}

\begin{figure*}
\centering
\includegraphics{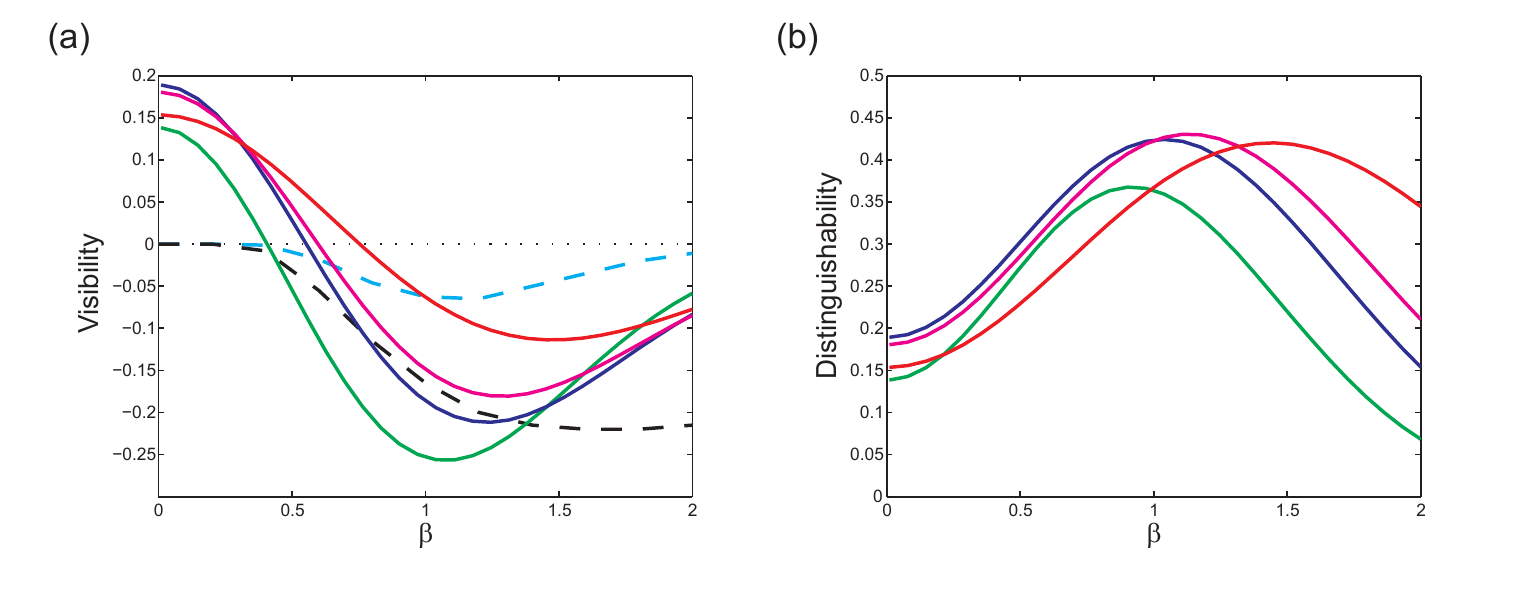}
\caption{Dependence of interference (visibility) $V(\beta)$ and distinguishability $D(\beta)$ on $\beta$ for squeezing of a single-photon state. Solid lines: green -- experimental single-photon state, blue -- experimental squeezed single-photon state ($\gamma = -0.38$), magenta -- experimental squeezed single-photon state ($\gamma = -0.50$), red -- experimental squeezed single-photon state ($\gamma = -0.72$).
Dashed lines in (a): black -- bound for mixtures of Gaussian states, cyan -- bound for mixtures of coherent states, dashed lines in (b): black -- single-photon state.
}\label{fig_suppl1}
\vspace{3\baselineskip}
\centering
\includegraphics{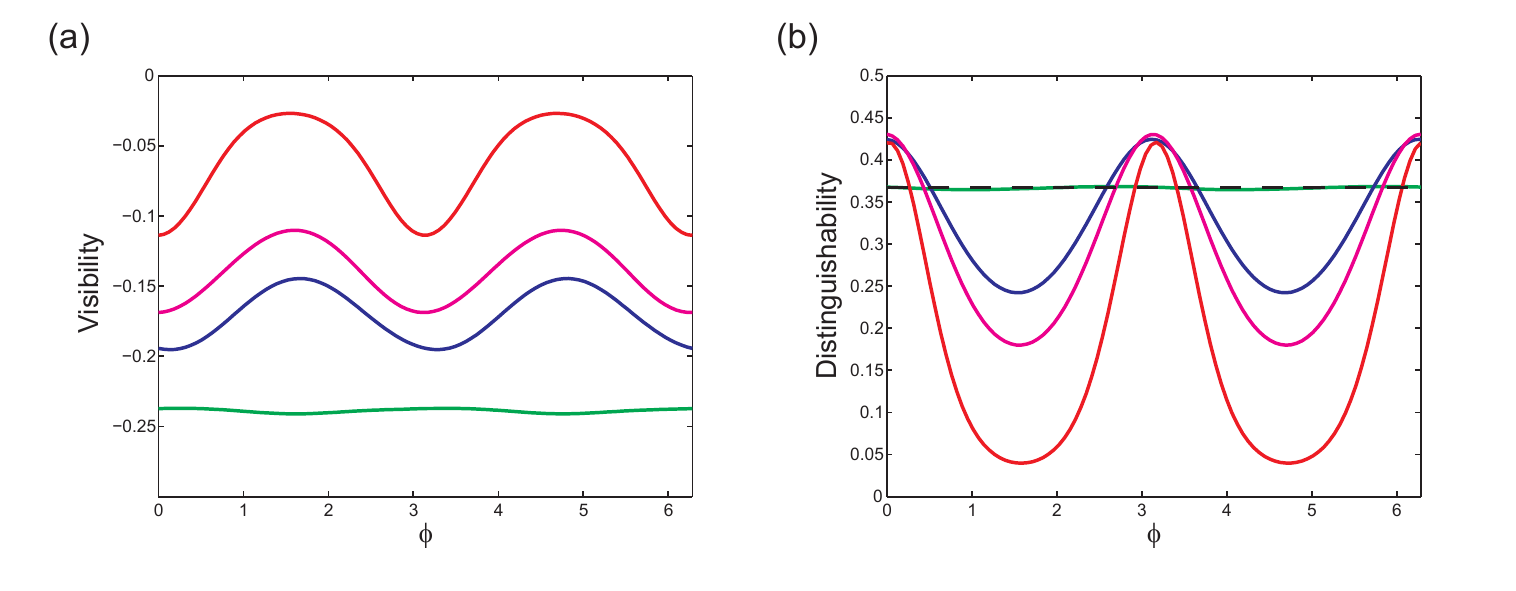}
\caption{Dependence of interference (visibility) $V(\beta_0 e^{i\phi})$ and distinguishability $D(\beta_0 e^{i\phi})$ on phase $\phi$ for squeezing of a single-photon state. For each curve, $\beta_0 \in \mathbb{R}$ was chosen as the point of the local extremum. (a): green -- experimental single-photon state ($\beta_0 = 1.11$), blue -- experimental squeezed single-photon state ($\gamma = -0.38, \beta = 1.26$), magenta -- experimental squeezed single-photon state ($\gamma = -0.50, \beta_0 = 1.31$), red -- experimental squeezed single-photon state ($\gamma = -0.72, \beta_0 = 1.45$). (b): green -- experimental single-photon state ($\beta_0 = 0.90$), blue -- experimental squeezed single-photon state ($\gamma = -0.38, \beta_0 = 1.04$), magenta -- experimental squeezed single-photon state ($\gamma = -0.50, \beta_0 = 1.11$), red -- experimental squeezed single-photon state ($\gamma = -0.72, \beta_0 = 1.45$). Dashed black -- single-photon state ($\beta_0 = 0.97$).
}\label{fig_suppl2}
\end{figure*}

The phase-insensitive superposition of coherent states present in a single-photon state, $\ket{1}\propto \int (\ket{\alpha e^{i\phi}} - \ket{-\alpha e^{i\phi}})d\phi$, is a typical manifestation of wave-particle duality -- even in the supposedly particle-like state $|1\rangle$, wave-like coherent states are present, but these form a continuous superposition with arbitrary phases and hence are not visible in the total state.
However, in a $\ket{1}$ $\rightarrow$ CSS conversion, the squeezer can break this rotational symmetry and amplify coherent states along a single axis, thus making them more distinguishable while simultaneously preserving their interference features. This is demonstrated in Suppl.~Fig.~\ref{fig_suppl1}, where we can see that squeezing not only increases distinguishability (which could as well be a consequence of Gaussian noise), but also shifts the maximum of $D(\beta)$ towards larger values of $\beta$, as one would expect from a larger CSS. At the same time, the minimum of $V(\beta)$ is also shifted, thus preserving the symmetry of an ideal CSS while remaining negative and distinct from a mixture of coherent states. The weakest squeezing operation ($\gamma=0.26$) even exhibits interference properties going beyond those of any Gaussian state. At the same time, in Suppl.~Fig.~\ref{fig_suppl2}, we can observe a phase modulation of $V(\beta)$ which emerges as a consequence of the phase-sensitive squeezing process.

For the opposite conversion, CSS $\rightarrow$ $\ket{1}$, the above criteria are ambiguous, because the transitions of $D(\alpha)$ and $V(\alpha)$ and their loss of the phase dependence can be as well explained by an incoherent dephasing
effect. However, by looking at the anti-correlation parameter $A$, which effectively serves as a measure of the `single-photon' quality of the state, we see that the values of this parameter change from $A_{\rm CSS} = 0.52$ to $A_{\rm SP} = 0.29$. Recalling that for an ideal single-photon state we have $A = 0$, this shows that the quality of  the single-photon state present in the total state is enhanced, even though the absolute single-photon fraction in the resulting mixed state is actually reduced.

\makeatletter
% \close@column@grid
\balancelastpage@sw{%
  \onecolumngrid
 }{}%
\twocolumngrid
\makeatother

\clearpage

\section{Supplemental Methods}

As a fundamental light source, we utilize a continuous-wave Ti:sapphire laser with a wavelength of $860$~nm.
Some part of the fundamental wave is converted to a second harmonic wave in order to pump the optical parametric oscillators (OPO-1 and 2).
As a non-classical input state, we generate single-photon states, or, alternatively, coherent-state superpositions (CSSs).
In Fig.~\ref{fig:setup} of the main text, the setup for creating these input states is visualized by an orange background. Compared to the setups of Refs.~\cite{JonasPhoton.optex} and \cite{WakuiCat.optex},
in the present scheme, we can switch between the two different non-classical input states.
In the case of a $\ket{1}$ input, non-degenerate photon pairs are generated in OPO-1 and one photon of every pair is then detected by an avalanche photo diode (APD).
A click of the APD at this arm heralds the presence of a single photon in the other arm.
The separating cavity (SC) behind OPO-1 allows for a
spatial separation of each pair of correlated photons.
The filtering cavities (FCs) extract a single frequency mode of OPO-1 from hundreds of modes
in every $590$~MHz of the hundreds GHz bandwidth of quasi-phase-matching,
where $590$~MHz is the free spectral range of OPO-1.
For the case of an initial CSS, OPO-1 generates degenerate photon pairs, i.e., squeezed vacuum states.
Every click of the APD then corresponds to the subtraction of a photon
%from the stream, namely, generations of
i.e., $\hat{a}\sq \ket{0}$.

Our measurement-based squeezer works as follows (visualized by a blue background in Fig.~\ref{fig:setup} of the main text) \cite{Filip05.pra}.
An ancillary $\hat{p}$-squeezed state and the input quantum state are coupled by a beam splitter, where the level of squeezing $\gamma$ can be adjusted by the beam-splitting ratio $T$ with $T = \exp(-2\gamma)$.
We demonstrate three different squeezing levels, $\gamma$ $=$ $0.26$, $0.37$, and $0.67$, corresponding to $T$ $=$ $0.59$, $0.48$, and $0.26$, respectively.
Then, one output of the beam splitter is measured with respect to the $\hat{x}$ quadrature and the outcome is fedforward to the $\hat{x}$ quadrature of the other output mode.
A negative squeezing parameter {\it e.g.\ }$\gamma = -0.26$ is achieved by $90^\circ$ rotations of the optical phases, corresponding to an $\hat{x}$-squeezed ancilla, a measurement with respect to $\hat{p}$, and a feedforward on $\hat{p}$.
The feedforward operation is performed as follows.
First, a phase modulation is added to an auxiliary beam by means of an electro-optic modulator (EOM), where the applied voltage is proportional to the measurement result.
Then, the auxiliary beam is coupled to the signal beam by a highly asymmetric beam splitter.
An optical delay of $12$~m in free space is utilized for a broadband synchronization of the feedforward \cite{CatTeleportation}.
Thus, we achieve broadband squeezing of up to $10$~MHz,
which is $300$ times broader than the conventional squeezing \cite{Yoshikawa07.pra}.
Note that the bandwidth limitation is that of the ancillary squeezed light.
The bandwidth is broad enough to operate a full $6$~MHz bandwidth of the heralded quantum states,
where $6$~MHz is the half-width-half-maximum of OPO-1.

In a preliminary experiment, the squeezing level of the ancilla squeezed vacuum state was $-6.8$ dB, while the antisqueezing level was $10.3$ dB.
The finite squeezing level of the ancilla causes some excess noise at the output of the squeezing gate, while the antisqueezing level of the ancilla does, in principle, not have any effect on the output, as will be explained below.
Phase fluctuations in the delay line can have the negative effect of degrading the squeezing level of the ancilla, because a part of the antisqueezed quadrature noise disturbs the conjugate squeezing quadrature.
However, in our experiment, the phase fluctuations are sufficiently suppressed by a feedback control to the level of a few degrees.

In addition, the beam pointing in the delay line is also actively controlled in the experiment via feedback, by which a significant decrease of the interference visibility can be prevented during the experiment.
The beam pointing control is implemented through a combination of a four-quarter photodiode (to detect misalignments) and a mirror mount attached to three piezoelectric actuators (to compensate misalignments).

As for the high-purity single-photon state with a single-photon component of $84\%$ used as the input state, among the remaining inefficiencies of $16\%$, the inefficiencies due to the detection process are estimated to be about $7\%$.
The measured optical propagation losses inside and outside the OPO are $4\%$ in total.
The contribution of fake signals from the APD corresponds to $1\%$.
Higher photon number components are about $2\%$.

% % added

%\subsection{Mathematical description}
%\label{ssec:matheq}

Finally, we shall briefly review the theoretical working principle of our measurement-based squeezing gate, as originally proposed in Ref.~\cite{Filip05.pra}.

An input state to be squeezed is represented by its quadrature operators $(\hat{x}_\text{in},\hat{p}_\text{in})$, while an ancilla squeezed vacuum state is represented by $(\hat{x}_\text{anc},\hat{p}_\text{anc})$.
A linear transformation of these quadrature operators in the Heisenberg representation uniquely represents the Gaussian gate operation.

First, the two optical states are combined by a beam splitter with a transmittance $T$, corresponding to the following transformations,
\begin{subequations}
\begin{align}
\hat{x}_\text{in}^\prime = & \sqrt{T}\hat{x}_\text{in} + \sqrt{1-T}\hat{x}_\text{anc}, \\
\hat{p}_\text{in}^\prime = & \sqrt{T}\hat{p}_\text{in} + \sqrt{1-T}\hat{p}_\text{anc}, \\
\hat{x}_\text{anc}^\prime = & - \sqrt{1-T}\hat{x}_\text{in} + \sqrt{T}\hat{x}_\text{anc}, \\
\hat{p}_\text{anc}^\prime = & - \sqrt{1-T}\hat{p}_\text{in} + \sqrt{T}\hat{p}_\text{anc},
\end{align}
\end{subequations}
where prime marks label those quadrature operators after the beam splitter interaction.

If we want to squeeze the $\hat{x}$ quadrature, we choose the ancilla state to be an $\hat{x}$-squeezed state, $\hat{x}_\text{anc} = \hat{x}_\text{vac}e^{-r}$, where $\hat{x}_\text{vac}$ is a quadrature operator of a vacuum state and $r\in\mathbb{R}$ is a squeezing parameter.
Next, we measure $\hat{p}_\text{anc}^\prime$, and then perform a displacement operation on the remaining state along the direction of the $\hat{p}$ quadrature by an amount proportional to the measurement outcome, corresponding to the transformation $\hat{x}_\text{out} = \hat{x}_\text{in}^\prime$ and $\hat{p}_\text{out} = \hat{p}_\text{in}^\prime - \sqrt{(1-T)/T}\hat{p}_\text{anc}^\prime$, leading to
\begin{subequations}
\begin{align}
\hat{x}_\text{out} =& \sqrt{T}\hat{x}_\text{in} + \sqrt{1-T}\hat{x}_\text{vac}e^{-r}, \\
\hat{p}_\text{out} =& \frac{1}{\sqrt{T}}\hat{p}_\text{in}.
\end{align}
\end{subequations}
As can be seen above, the ancilla's antisqueezed noise is canceled by the feedforward operation, and only the squeezed quadrature of the ancilla remains as an extra term.
In the limit $r\to\infty$, this extra term becomes negligibly small, and the overall transformation approaches an ideal squeezing operation $\hat{S}(\gamma)=e^{\gamma(\hat{a}^{\dagger2}-\hat{a}^2)/2}$ with $\gamma= \ln\sqrt{T} \le 0$.

Similarly, if we want to squeeze the $\hat{p}$ quadrature, we choose the ancilla state to be a $\hat{p}$-squeezed state, $\hat{p}_\text{anc} = \hat{p}_\text{vac}e^{-r}$, where $\hat{p}_\text{vac}$ is a quadrature operator of a vacuum state.
This time we measure $\hat{x}_\text{anc}^\prime$ and then perform a displacement operation on the remaining state along the direction of the $\hat{x}$ quadrature, corresponding to the transformation $\hat{x}_\text{out} = \hat{x}_\text{in}^\prime - \sqrt{(1-T)/T}\hat{x}_\text{anc}^\prime$ and $\hat{p}_\text{out} = \hat{p}_\text{in}^\prime$, this time resulting in
\begin{subequations}
\begin{align}
\hat{x}_\text{out} =& \frac{1}{\sqrt{T}}\hat{x}_\text{in}, \\
\hat{p}_\text{out} =& \sqrt{T}\hat{p}_\text{in} + \sqrt{1-T}\hat{p}_\text{vac}e^{-r}.
\end{align}
\end{subequations}
This transformation approaches an ideal squeezing operation $\hat{S}(\gamma)$ with $\gamma= -\ln\sqrt{T} \ge 0$ in the limit $r\to\infty$.

All the elements of this squeezing gate, i.e., the creation of an ancilla squeezed vacuum state, the beam splitter interaction, the homodyne detection, and the feedforward displacement operation, are each realizable in an unconditional and deterministic fashion, and hence the squeezing gate as a whole also works unconditionally and deterministically.
The only imperfection is the excess noise in the squeezed quadrature due to the finite ancilla squeezing.


\begin{thebibliography}{99}


\bibitem{KLM}
E.\ Knill, R.\ Laflamme, and G.\ J.\ Milburn,
% \articletitle{A scheme for efficient quantum computation with linear optics}
\nature {\bf 409}, 46--52 (2001).

\bibitem{Ionicioiu11}
R.\ Ionicioiu and D.\ R.\ Terno,
% \articletitle{Proposal for a quantum delayed-choice experiment}
\prl {\bf 107}, 230406 (2011).

\bibitem{Glauber63}
R.\ J.\ Glauber,
% \articletitle{Coherent and incoherent states of the radiation field}
\prev {\bf 131}, 2766--2788 (1963).

%\bibitem{DLCZ}
%L.\ M.\ Duan, M.\ D.\ Lukin, J.\ I.\ Cirac, and P.\ Zoller,
%\articletitle{Long-distance quantum communication with atomic ensembles and linear optics}
%\nature {\bf 414}, 413--418 (2001).

\bibitem{CatTeleportation}
N.\ Lee, H.\ Benichi, Y.\ Takeno, S.\ Takeda, J.\ Webb, E.\ Huntington, and A.\ Furusawa,
% \articletitle{Teleportation of nonclassical wave packets of light}
\science {\bf 332}, 330--333 (2011).

\bibitem{DRQubitTeleportation}
S.\ Takeda, T.\ Mizuta, M.\ Fuwa, P.\ van Loock, and A.\ Furusawa,
% \articletitle{Deterministic quantum teleportation of photonic quantum bits by a hybrid technique}
\nature {\bf 500}, 315--318 (2013).

\bibitem{hybridbook}
A.\ Furusawa and P.\ van Loock,
{\it Quantum Teleportation and Entanglement: A Hybrid Approach to Optical Quantum Information Processing} (Wiley VCH, Berlin, 2011).

%\bibitem{LloydbBraunstein}
%S.\ Lloyd and S.\ L.\ Braunstein,
%\articletitle{Quantum computation over continuous variables}
%\prl {\bf 82}, 1784 (1999).

%\bibitem{Ourjoumtsev06}
%A.\ Ourjoumtsev, R.\ Tualle-Brouri, J.\ Laurat, and P.\ Grangier,
%\articletitle{Generating optical Schr\"{o}dinger kittens for quantum information processing}
%\science {\bf 312}, 83--86 (2006).

%\bibitem{Jonas06}
%J.\ S.\ Neergaard-Nielsen, B.\ Melholt Nielsen, C.\ Hettich, K.\ M\/{o}lmer, and E.\ S.\ Polzik,
%\articletitle{Generation of a superposition of odd photon number states for quantum information networks}
%\prl 97, 083604 (2006).

\bibitem{HistoricalSqueezingRef}
D.\ F.\ Walls and G.\ J.\ Milburn,
{\it Quantum Optics} (Springer-Verlag, Berlin Heidelberg New York, 1994).

\bibitem{RMP_Braunstein}
S.\ L.\ Braunstein and P.\ van Loock,
% \articletitle{Quantum information with continuous variables}
\rmp {\bf 77}, 513--577 (2005).

%\bibitem{HongOuMandel}
%C.\ K.\ Hong, Z.\ Y.\ Ou, and L.\ Mandel,
%\articletitle{Measurement of subpicosecond time intervals between two photons by interference}
%\prl {\bf 59}, 2044--2046 (1987).

%\bibitem{StandardSqueezingExperiments}
%Y.\ Takeno, M.\ Yukawa, H.\ Yonezawa, and A.\ Furusawa,
%\articletitle{Observation of $-9$~dB quadrature squeezing with improvement of phase stability in homodyne measurement}
%\optexp {\bf 15}, 4321--4327 (2007).

\bibitem{RMP_Weedbrook}
C.\ Weedbrook, S.\ Pirandola, R.\ Garc\'ia-Patr\'on, N.\ J.\ Cerf, T.\ C.\ Ralph, J.\ H.\ Shapiro, and S.\ Lloyd,
% \articletitle{Gaussian quantum information}
\rmp {\bf 84}, 621--669 (2012).

\bibitem{irreducible}
S.\ L.\ Braunstein,
% \articletitle{Squeezing as an irreducible resource}
\pra {\bf 71}, 055801 (2005).

\bibitem{x3gate}
P.\ Marek, R.\ Filip, and A.\ Furusawa,
% \articletitle{Deterministic implementation of weak quantum cubic nonlinearity}
\pra {\bf 84}, 053802 (2011).

\bibitem{Yoshikawa07.pra}
J.\ Yoshikawa, T.\ Hayashi, T.\ Akiyama, N.\ Takei, A.\ Huck, U.\ L.\ Andersen, and A.\ Furusawa,
% \articletitle{Demonstration of deterministic and high fidelity squeezing of quantum information}
\pra {\bf 76}, 060301(R) (2007).

\bibitem{Yoshikawa08.prl}
J.\ Yoshikawa, Y.\ Miwa, A.\ Huck, U.\ L.\ Andersen, P.\ van Loock, and A.\ Furusawa,
% \articletitle{Demonstration of a quantum nondemolition sum gate}
\prl {\bf 101}, 250501 (2008).

\bibitem{Yoshikawa11.pra}
J.\ Yoshikawa, Y.\ Miwa, R.\ Filip, and A.\ Furusawa,
% \articletitle{Demonstration of a reversible phase-insensitive optical amplifier}
\pra {\bf 83}, 052307 (2011).

\bibitem{Supplemental}
See the Supplemental Material for additional evaluation of the conversion and for further experimental methods.

\bibitem{CSSfromPhotonEx}
A.\ Ourjoumtsev, H.\ Jeong, R.\ Tualle-Brouri, and P.\ Grangier,
% \articletitle{Generation of optical `Schr\"{o}dinger cats' from photon number states}
\nature {\bf 448}, 784--786 (2007).

\bibitem{CSS_comp}
T.\ C.\ Ralph, A.\ Gilchrist, G.\ J.\ Milburn, W.\ J.\ Munro, and S.\ Glancy,
% \articletitle{Quantum computation with optical coherent states}
\pra {\bf 68}, 042319 (2003).

\bibitem{Yoshikawa13}
J.\ Yoshikawa, K.\ Makino, S.\ Kurata, P.\ van Loock, and A.\ Furusawa,
% \articletitle{Creation, storage, and on-demand release of optical quantum states with a negative Wigner function}
\prx {\bf 3}, 041028 (2013).

\bibitem{Takeoka}
M.\ Takeoka and M.\ Sasaki,
% \articletitle{Discrimination of the binary coherent signal: Gaussian-operation limit and simple non-Gaussian near-optimal receivers}
\pra {\bf 78}, 022320 (2008).

\bibitem{Menzies09.pra}
D.\ Menzies and R.\ Filip,
% \articletitle{Gaussian-optimized preparation of non-Gaussian pure states}
\pra {\bf 79}, 012313 (2009).

\bibitem{Filip13.pra}
R.\ Filip,
% \articletitle{Gaussian quantum adaptation of non-Gaussian states for a lossy channel}
\pra {\bf 87}, 042308 (2013).

\bibitem{Interface}
P.\ Marek and R.\ Filip,
% \articletitle{Noise-resilient quantum interface based on quantum nondemolition interactions}
\pra {\bf 81}, 042325 (2010).

\bibitem{Zaidi}
H.\ Zaidi and P.\ van Loock,
% \articletitle{Beating the one-half limit of ancilla-free linear optics Bell measurements}
\prl {\bf 110}, 260501 (2013).

\bibitem{AkiraDistillationExperiment}
H.\ Takahashi, J.\ S.\ Neergaard-Nielsen, M.\ Takeuchi, M.\ Takeoka, K.\ Hayasaka, A.\ Furusawa, and M.\ Sasaki,
% \articletitle{Entanglement distillation from Gaussian input states}
\natphoton {\bf 4}, 178--181 (2010).

\bibitem{TheorySqueezingHelps}
S.\ Zhang and P.\ van Loock,
% \articletitle{Local Gaussian operations can enhance continuous-variable entanglement distillation}
\pra {\bf 84}, 062309 (2011).

\bibitem{JonasPhoton.optex}
J.\ S.\ Neergaard-Nielsen, B.\ Melholt Nielsen, H.\ Takahashi, A.\ I.\ Vistnes, and E.\ S.\ Polzik,
% \articletitle{High purity bright single photon source}
\optexp {\bf 15}, 7940--7949 (2007).

\bibitem{WakuiCat.optex}
K.\ Wakui, H.\ Takahashi, A.\ Furusawa, and M.\ Sasaki,
% \articletitle{Photon subtracted squeezed states generated with periodically poled KTiOPO$_4$}
\optexp {\bf 15}, 3568--3574 (2007).

\bibitem{Filip05.pra}
R.\ Filip, P.\ Marek, and U.\ L.\ Andersen,
% \articletitle{Measurement-induced continuous-variable quantum interactions}
\pra {\bf 71}, 042308 (2005).

\bibitem{Andersen04.prl}
U.\ L.\ Andersen, O.\ Gl\"{o}ckl, S.\ Lorenz, G.\ Leuchs, and R.\ Filip,
% \articletitle{Experimental demonstration of continuous variable quantum erasing}
\prl {\bf 93}, 100403 (2004).

\bibitem{Lvovsky04}
A.\ I.\ Lvovsky,
% \articletitle{Iterative maximum-likelihood reconstruction in quantum homodyne tomography}
\job {\bf 6}, S556--S559 (2004).

\bibitem{Lvovsky09.rmp}
A.\ I.\ Lvovsky and M.\ G.\ Raymer,
% \articletitle{Continuous-variable optical quantum-state tomography}
\rmp {\bf 81}, 299--332 (2009).

%\bibitem{Bartlett02.prl}
%S.\ D.\ Bartlett, B.\ C.\ Sanders, S.\ L.\ Braunstein, and K.\ Nemoto,
%\articletitle{Efficient classical simulation of continuous variable quantum information processes}
%\prl {\bf 88}, 097904 (2002).

\bibitem{CSSfromPhoton}
A.\ P.\ Lund, H.\ Jeong, T.\ C.\ Ralph, and M.\ S.\ Kim,
% \articletitle{Conditional production of superpositions of coherent states with inefficient photon detection}
\pra {\bf 70}, 020101(R) (2004).

\bibitem{anti-correlation}
P.\ Grangier, G.\ Roger, and A.\ Aspect,
% \articletitle{Experimental evidence for a photon anticorrelation effect on a beam splitter: a new light on single-photon interference}
\epl {\bf 1}, 173--179 (1986).


\end{thebibliography}
\end{document}